\documentclass[a4paper,fleqn]{cas-sc}

\def\tsc#1{\csdef{#1}{\textsc{\lowercase{#1}}\xspace}}
\tsc{WGM}
\tsc{QE}
\tsc{EP}
\tsc{PMS}
\tsc{BEC}
\tsc{DE}

\begin{document}
\let\WriteBookmarks\relax
\def\floatpagepagefraction{1}
\def\textpagefraction{.001}
\shorttitle{Rotating Magnetocaloric Effect in Sintered La(Fe,Mn,Si)$_{13}$H$_z$ Plates}
\shortauthors{R. Almeida et~al.}

\title [mode = title]{Rotating Magnetocaloric Effect in Sintered La(Fe,Mn,Si)$_{13}$H$_z$ Plates}                      

\author[1]{Rafael Almeida}
\credit{Conceptualization, Software, Formal analysis, Investigation, Data Curation, Writing – Original Draft, Writing – Review \& Editing, Visualization}

\author[1]{Tomás Ventura}
\credit{Formal analysis, Investigation, Data Curation, Writing – Original Draft, Writing – Review \& Editing}

\author[1]{Ricardo Moura Costa Pinto}
\credit{Investigation, Writing – Review \& Editing}

\author[1]{João Oliveira Silva}
\credit{Investigation, Writing – Review \& Editing}

\author[2]{Konrad Loewe}
\credit{Resources, Investigation, Formal analysis, Resources, Writing – Review \& Editing}

\author[3]{Rodrigo Kiefe}
\credit{Methodology, Software, Validation, Writing – Review \& Editing}

\author[3]{João Sequeira Amaral}
\credit{Conceptualization, Investigation, Writing – Review \& Editing, Funding Acquisition}

\author[1]{João Pedro Araújo}
\credit{Writing – Review \& Editing, Supervision, Funding acquisition}
        
\author[1]{João Horta Belo}
\cormark[1]
\credit{Conceptualization, Investigation, Writing – Review \& Editing, Supervision, Funding acquisition, Project administration}

\cortext[cor1]{Corresponding author: jbelo@fc.up.pt}

\affiliation[1]{organization={IFIMUP – Institute of Physics for Advanced Materials, Nanotechnology and Photonics},
                country={Department of Physics and Astronomy, Faculty of Sciences of University of Porto, rua do Campo Alegre, s/n, 4169-007 Porto, Portugal}}
\affiliation[2]{organization={Vacuumschmelze GmbH \& Co. KG},
                country={63450 Hanau, Germany}}
\affiliation[3]{organization={Physics Department and CICECO — Aveiro Institute of Materials},
                country={University of Aveiro, 3810-193 Aveiro, Portugal}}

\begin{abstract}
La-Fe-Si-based alloys are among the most application-ready magnetocaloric materials for room-temperature magnetic refrigeration. Powder metallurgy methods have been previously demonstrated to successfully produce structures with sub-mm features for magnetic refrigerators in a scalable method. In this work, we explore the rotating magnetocaloric effect (RMCE) present in a 0.27 mm thin plate of sintered and hydrogenated La(Fe,Mn,Si)$_{13}$. The high aspect ratio ($\sim$50) of the thin plate leads to an anisotropic magnetocaloric effect (MCE), dependent on the relative orientation of the external magnetic field, and an RMCE when the external field is rotated. We find a maximum rotating adiabatic temperature change ($\Delta T_{ad}^{rot}$) of 1.17 K with the rotation of a 1 T magnetic field and 1.12 K when rotating a 0.6 T magnetic field, a reduction of only 4\% for a 40\% reduction in applied field strength. Magnetostatic computations revealed a considerable rotating isothermal entropy change ($\Delta S_{iso}^{rot}$), comparable to the conventional MCE of Gd for similar fields, reaching 3.97 J K$^{-1}$ kg$^{-1}$ for 1 T and 3.68 J K$^{-1}$ kg$^{-1}$ for 0.6 T (7\% reduction), highlighting La-Fe-Mn-Si alloys as high potential candidates for a magnetic refrigerator based on the RMCE utilizing relatively low external magnetic field amplitudes, such as 0.6 T. 
\end{abstract}


\maketitle

\section{Introduction}

\subsection*{Magnetocaloric refrigeration and La-Fe-Si alloys}

Economic development and global population growth will significantly increase demand for heat pumping technologies in the coming decades \cite{IEA2018,IIR2025}. At the same time, the use of high global warming potential (GWP) refrigerants in vapor compression is being phased down due to environmental concerns, renewing interest in the use of natural refrigerants with low GWP such as CO$_2$, and fourth generation refrigerants, hydrofluoroolefins (HFOs) \cite{VUPPALADADIYAM2022,SONG2022}. Besides innovation in vapor compression, in recent years there have been considerable efforts in the development of alternative cooling technologies such as caloric (magnetocaloric \cite{Zhang2024}, elastocaloric \cite{CHEN2021102706}, barocaloric \cite{CIRILLO2022101380}, and electrocaloric \cite{SHI20191200}), thermoacoustic \cite{XIAO2024101815}, adsorption \cite{CHAUHAN2022111808}, and evaporative cooling \cite{ZEOLI2025136257}.

Within caloric technologies, magnetocaloric cooling is at the most advanced stages, with several prototypes and numerical models showing promising device performance \cite{Johra_Bahl_2022,johra2025_tech}. The phenomenon behind magnetocaloric cooling is the magnetocaloric effect (MCE), exhibited by magnetocaloric materials (MCMs) within the vicinity of a magnetic phase transition. \cite{Franco2018}. 

While some challenges remain \cite{ZHANG2021110933,Klinar2024}, significant advances in magnetocaloric cooling devices have been made possible by discoveries and improvements of MCMs over the past decades. Pure Gd, with a magnetic second order phase transition (SOPT) has functioned as the baseline MCM for room temperature applications due to its considerable MCE over a wide temperature range and predictable behaviour, but there are now some material families with giant magnetocaloric effect (GMCE) and negligible hysteresis which have lead to some better performing devices at room temperature \cite{APREA2018370,JACOBS201484}. 

Among many MCMs, the alloys based on La-Fe-Si and Mn-Fe-P-Si stand out due to the high tunability of their Curie temperature through slight changes of element fractions, allowing the design of layered active magnetic regenerators (AMR) \cite{Kitanovski2015,APREA201197,TUSEK2014117,Tomc2025layering}, and lack of heavy rare-earth elements, leading to lower cost \cite{Gottschall2019}. Furthermore, it is possible to manufacture fully dense 3D structures of La-Fe-Si-based alloys with sub-milimeter features through powder metallurgy, which is a highly scalable technique \cite{Katter2008, LIANG2021116519}. Such structures optimize heat transfer and low pressure drop within AMRs, and have recently been reported to perform stably in realistic magnetocaloric heat pump conditions \cite{LIONTE202143,Weiss2025}.

The outstanding magnetocaloric properties of the La-Fe-Si alloys were initially found between 180 K and 210 K in the La(Fe$_x$Si$_{1-x}$)$_{13}$ composition with $0.87\leq x\leq0.90$ \cite{Hu2001}. Soon after, it was found that hydrogenation could increase the Curie temperature of the alloys to 280-340 K without negatively impacting magnetocaloric properties \cite{Fujieda2002,Fujita2003}. It was also reported that partially substituting Fe with Co could increase T$_C$ beyond room temperature without hydrogenation while reducing hysteresis, at the "cost" of gradually decreasing the intensity of the MCE \cite{Hu2002CoSi,Katter2008}. By combining partial hydrogenation and Co substitution, one could tune the T$_C$ to room temperature while maintaining a considerable MCE \cite{Katter2008DDMC}. Finally, it was found that by substituting Fe for Mn, the T$_C$ could be tuned without significantly impacting the MCE, which made possible achieving T$_C$ at room temperature with complete hydrogenation, preventing heterogeneity in samples due to hydrogen migration \cite{Barcza2011}. More detailed accounts of these initial developments may be found elsewhere \cite{LIU2012584,Basso2015}. 

In this work, we present the demagnetizing field-based rotating magnetocaloric effect (RMCE) in a fully hydrogenated La(Fe,Mn,Si)$_{13}$ sample with 0.27 mm thickness produced through powder metallurgy \cite{Katter2008, Barcza2011}.

\subsection*{The demagnetizing field-based RMCE}

The demagnetizing effect is a universal phenomenon in magnetic samples with finite sizes, in which the magnetization of the material generates a magnetic field with opposing orientation within the sample volume - the demagnetizing field. With the exception of ellipsoidal samples \cite{Cronemeyer1991}, its orientation and amplitude vary in space, but a scalar approximation is useful and accurate in many simple shapes. This approximation can be expressed as:

\begin{equation}
\label{eq:1}
    H_{int} = H_{ext}-DM,
\end{equation}
where $H_{int}$ and $H_{ext}$ are the (average) magnetic field strength respectively within and outside the sample volume, and $-DM$ is the demagnetizing field strength, proportional to the sample's magnetization, $M$, and the shape-dependent demagnetizing factor, $D$. In SI units, $D$ can vary between 0 (when the magnetic field is oriented along a dimension of the sample much longer than the others, such the in-plane orientation in a thin film) and 1 (when the magnetic field is oriented along a dimension of the sample much shorter than the others, such as the off-plane orientation in a thin film).

While the impact of the demagnetizing effect in the MCE is well documented \cite{Kuzmin2011,Christensen_2011,Smith2012}, it was very rarely considered as a potential source of RMCE achievable by rotating an asymmetrically shaped sample, changing its demagnetizing factor and thus the internal magnetic field to which the material is exposed \cite{Barclay1984,Mansanares2013, Badosa2023}. Only more recently has its interesting external magnetic-field dependence been detailed, suggesting that a considerable RMCE could be obtained in any MCM in relatively low fields \cite{Almeida2024}.

While there are numerous ways of designing magnetic field sources within magnetic refrigeration devices \cite{TOMC2023157}, the most often implemented architecture for devices operating at room temperature is based on permanent magnets, as opposed to electromagnets or superconducting magnets. Due to their high price, Nd-Fe-B magnets are often the most expensive components within these devices, as their volume quickly increases with the desired maximum magnetic field strength \cite{RUSSEK20061366}. The RMCE enables a simpler, more compact design as it does not require alternating the MCM between regions with high and low magnetic field. Instead, the MCM can remain in a region of constant magnetic field intensity at all times, only requiring that the magnetic field rotate, potentially leading to a more economically viable device design \cite{BARCLAY1980467,FERNANDES2025134,FERNANDES2025272}.

\begin{figure}
    \centering
    \includegraphics[width=160mm]{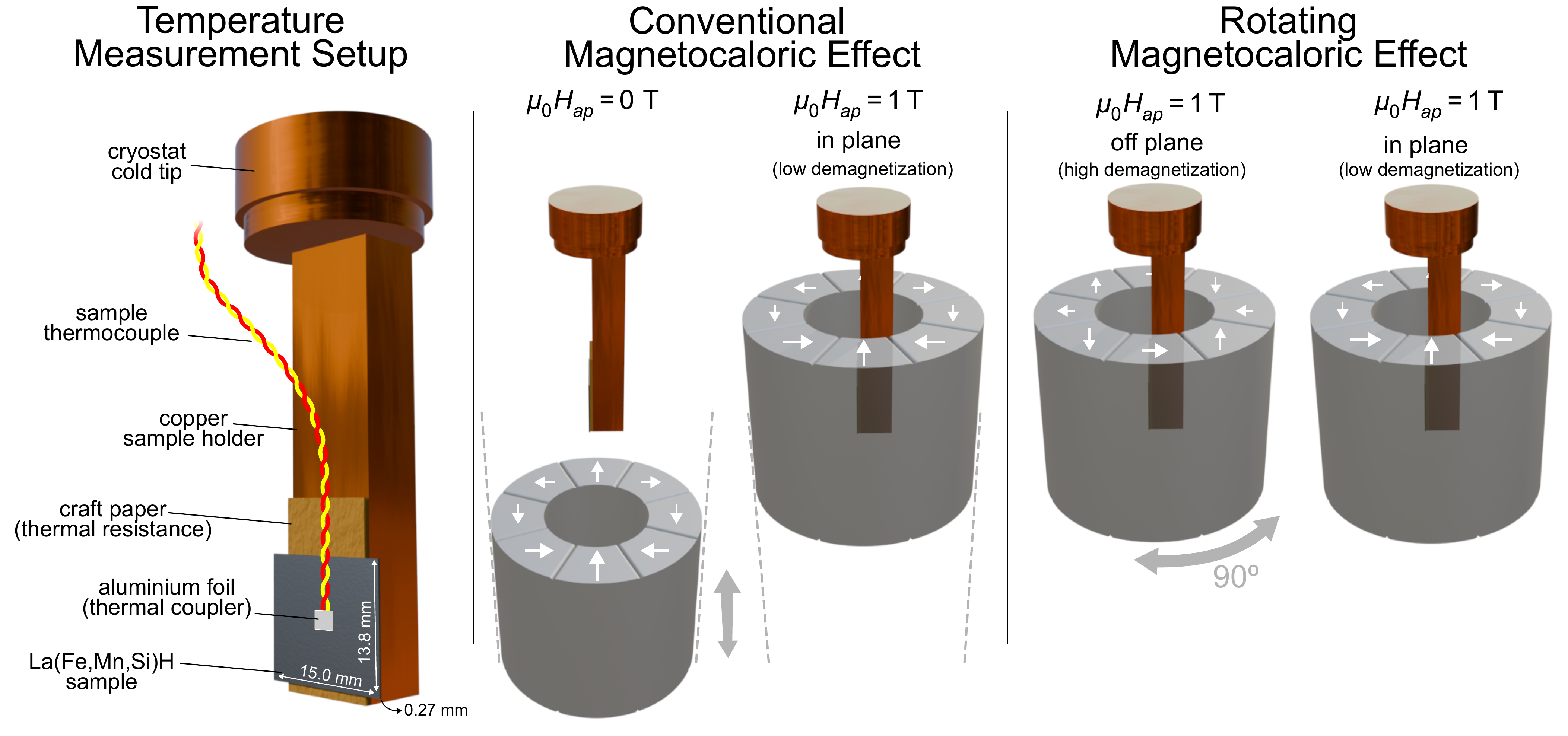}
    \caption{Schematic view of the temperature measurement setup (left), and procedure to measure the conventional (center) and rotating magnetocaloric effect (right) with a Halbach cylinder providing 1 T.}
    \label{fig:1}
\end{figure}

\section{Experimental Details}

A fully dense La-Fe-Mn-Si sample was manufactured through powder metallurgy as described in \cite{Katter2008}. After sintering, a thin (0.27 mm) plate was cut and fully hydrogenated. The adiabatic temperature change, $\Delta T_{ad}$, and rotating adiabatic temperature change, $\Delta T_{ad}^{rot}$, exhibited when applying or rotating an external magnetic field with constant intensity (0.6 T or 1.0 T), were directly measured with a type-K thermocouple of 75 $\mathrm{\mu m}$-wire diameter. One effective method for installing a thermocouple is to sandwich the sensing tip between two halves of the sample \cite{Salazar2023}. In this work, to improve thermal coupling without compromising the sample's high aspect ratio (which enhances the demagnetizing field-based RMCE), a small piece of aluminium foil was used to remove excess adhesive (GE-varnish) and fix the thermocouple tip to the sample, with measurements made under vacuum ($P<10^{-4}$ mbar). Two layers of craft paper were used as a thermal resistance to slow the heat transfer to the copper sample holder, further improving adiabatic conditions. The baseline temperature was controlled by installing the sample holder in a closed-loop cryostat with a heater near the cold tip. The external magnetic field was applied/rotated in under a second by displacing or rotating a Halbach cylinder providing 1.0 T, and another for providing 0.6 T. A schematic of the temperature measurement setup and side view is shown in figure \ref{fig:1}.

Magnetization measurements were performed in a smaller piece (13.8 mg) of the same original sample using an MPMS3 SQUID magnetometer by Quantum Design.

To corroborate the direct measurements of $\Delta T_{ad}^{rot}$ and evaluate the rotating isothermal entropy change, $\Delta S_{iso}^{rot}$, magnetostatic simulations were performed using the experimental magnetization as input. These simulations give us the spatially dependent internal magnetic field corresponding to the experimentally measured sample's shape at the different combinations of: temperature; external magnetic field intensities; and orientations of the magnetic field. Using a mesh with over 200k elements, the average internal fields were obtained for the in-plane (low demagnetization) and off-plane (high demagnetization) orientations up to 2.0 T. The simulations were done using the open access software \textit{FEMCE} \cite{Kiefe2025}. For cross validation, the computations were repeated using 1) COMSOL Multiphysics and 2) numerically solving the scalar approximation shown in equation \ref{eq:1}, yielding equivalent results in all cases (less than 0.006 T difference in average internal field in-plane, and less than 0.016 T off-plane).

\section{Results and Analysis}

\begin{figure}
    \centering
    \includegraphics[width=70mm]{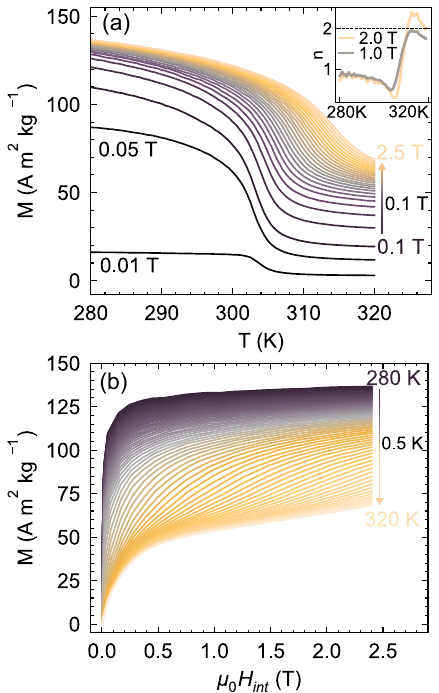}
    \caption{Magnetization versus temperature and magnetic field (a) as-measured, in iso-field curves obtained on heating for different external magnetic field intensities, and (b) after demagnetization correction, versus internal magnetic field, shown for temperatures between 280 K and 320 K with a step of 0.5 K.}
    \label{fig:2}
\end{figure}

The magnetic response of the sample is shown in figure \ref{fig:2}(a) as measured in iso-field conditions on heating, and in figure \ref{fig:2}(b), the same data is plotted as a function of $H_{int}$, after performing demagnetization correction using equation \ref{eq:1}. The material exhibits a ferromagnetic to paramagnetic phase transition at 302 K, which gradually shifts towards higher temperatures as the magnetic field intensity increases, becoming less sharp. 

The T$_C$ of the La(Fe$_{x}$,Mn$_y$,Si$_{1-x-y}$)$_{13}$ alloys has been previously studied as a function of Mn content, $y$ \cite{Basso2015}. A T$_C$ of 302 K corresponds to a 2 at\% substitution of Fe, with $x=0.88$ and $y=0.02$ \cite{Basso2015}. This is compatible with a SEM-EDX scan of our sample's main (1-13) phase, which yielded $x=0.883 \pm 0.003$ and $y=0.018 \pm 0.002$.

The magnetization data also reveals a strong ferromagnetic (FM) response, even above the phase transition temperature. This is attributable to the secondary $\alpha$-Fe phase, whose Curie temperature is above 1000 K \cite{Danan1968}. Assuming an ideal Curie-Weiss response at 380 K (constant susceptibility) of the paramagnetic (PM) 1-13 primary phase and a Fe saturation magnetization of 222 A m$^2$ kg$^{-1}$ \cite{Danan1968}, we estimated a 21.4 wt\% content of $\alpha$-Fe (see figure \ref{A1} of the appendix). Surface analysis of the sample using backscattered electron imaging in SEM estimated a slightly higher 28.0 vol.\% $\alpha$-Fe content (see figure \ref{A2} of the appendix). This secondary phase does not contribute to the MCE at room temperature, but enhances the alloy's mechanical stability, preventing crack propagation in the soft 1–13 primary phase, and thereby increasing its durability over many cycles across the phase transition \cite{SHAO2017506}.

The sample exhibited low thermal hysteresis in magnetization (a 0.6 K shift near T$_C$ between heating and cooling curves with an external field of 0.01 T), indicating "barely FOPT" behavior. Indeed, this is one of the practical advantages of these alloys, which display the advantage of FOPT while maintaining very low hysteresis. One quantitative criterion to distinguish SOPT and FOPT materials is the exponent $n$ of the internal magnetic field strength dependence of the isothermal entropy change, $\Delta S_{iso}$ \cite{Law2018}:

\begin{equation}
    \Delta S_{iso} \propto H^n,
\end{equation}
which can be computed through a polynomial fit for a certain temperature or locally at any temperature and field through: 

\begin{equation}
\label{eq:2}
    n\,(T,H_{int}) = \frac{\partial \ln |\Delta S_{iso}|}{\partial \ln H_{int}},
\end{equation}
where $\Delta S_{iso}$ is defined by 

\begin{equation}
    \label{eq:3}
    \Delta S_{iso}(T,H_{int,f}) = \int _0 ^{H_{int,f}} \frac{\partial M(T,H_{int})}{\partial T} dH_{int}.
\end{equation}

The exponent $n$ is shown inset in figure \ref{fig:2}(a), and corroborates this "barely" FOPT nature, as $n$ approaches 2 for 1.0 T (SOPT behavior) and slightly surpasses 2 for 2.0 T (FOPT behavior) at temperatures above T$_C$. The $n=2$ threshold is a distinguishing feature of materials with FOPT ($n>2$) and SOPT ($n<2$) \cite{Law2018}. Nevertheless, the slight hysteresis motivated measuring the full $M(T,H)$ dataset under consistent conditions (iso-field while heating) as opposed to mixing data acquired under cooling and heating \cite{Bez2018}. 

\begin{figure}
    \centering
    \includegraphics[width=160mm]{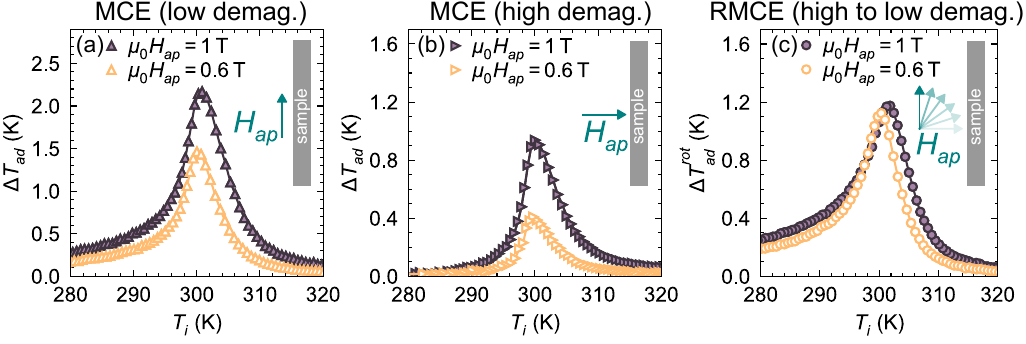}
    \caption{Directly measured $\Delta T_{ad}$ when applying an external magnetic field (a) along the low demagnetizing factor orientation and (b) along the high demagnetizing factor orientation of the sample, as shown in the inset schematic. (c) Directly measured $\Delta T_{ad}^{rot}$ when rotating the external magnetic field. The relative orientation of the sample (not to scale) and external magnetic field is schematically shown.}
    \label{fig:3}
\end{figure}

The direct $\Delta T_{ad}$ and $\Delta T_{ad} ^{rot}$ measurements of the conventional MCE and RMCE are shown in figure \ref{fig:3}. Under 1.0 T applied field, the maximum $\Delta T_{ad}$ observed was 2.17 K, when applying the external magnetic field in the low demagnetization orientation (parallel to the sample surface and along the longest sample dimension). While this value is in the higher end of what is reversibly achievable with 1.0 T \cite{Gottschall2019}, it is nevertheless 25\% lower than previously achieved in a similar sample (2.89 K) \cite{Basso2015}. This discrepancy may be due to the high (21.5 wt\%) $\alpha$-Fe content in our sample, which is not specified in the reference data. 

If the 1.0 T field is applied in the high demagnetization orientation, the maximum $\Delta T_{ad}$ is drastically reduced to less than half, 0.94 K (figure \ref{fig:3}(b)). In figure \ref{fig:3}(c), we can see that the maximum $\Delta T_{ad}^{rot}$ obtained when rotating the 1.0 T applied field is 1.17 K, larger than the conventional MCE obtained in the high demagnetization orientation. This is in fact true even beyond the maximum: especially at lower temperatures, for both measured magnetic field intensities, the RMCE is higher than the MCE measured on the high demagnetization orientation. 

The demagnetizing field-based RMCE may be estimated by subtracting the conventional MCE obtained in the two magnetic field orientations, as pointed out previously \cite{Almeida2024}. Indeed, the difference between the maximum $\Delta T_{ad}$ obtained in the two orientations, 1.24 K, is within 6\% of the maximum $\Delta T_{ad}^{rot}$ measured with the same field.

The $\Delta T_{ad}$ measured for 0.6 T are, unsurprisingly, lower than those obtained for 1.0 T, however, not in the same proportion for the low and high demagnetization orientations. Consequently, the relative intensity of the RMCE with respect to the MCE at the high demagnetization orientation is further accentuated. To better analyze this differences, it is useful to define the exponent $p$ of the magnetic field dependence of the peak $\Delta T_{ad}$:

\begin{equation}
    \Delta T_{ad}^{pk} \propto H^p.
\end{equation}

In the low demagnetization orientation, we obtain $p=0.77$, which is a somewhat stronger dependence than observed for other materials, such as $p=0.67$ \cite{Kuzmin2011} (Gd, LaFe$_{11.2}$Si$_{1.8}$, Nd$_2$Fe$_{17}$), or $p=0.70$ \cite{Franco2009} (Gd). On the other hand, for the high demagnetization orientation, we observed a much higher $p=1.63$. This means the maximum $\Delta T_{ad}$ obtained for 0.6 T is 67\% of the value obtained for 1.0 T in the low demagnetization orientation (reduced from 2.17 K to 1.46 K) and only 44\% in the high demagnetization orientation (reduced from 0.94 K to 0.41 K). These results are summarized in table 1.

\begin{table}[width=.9\linewidth,cols=5,pos=h]
\label{table:1}
\caption{Summary of the direct measurements of $\Delta T_{ad}$ for both applied magnetic field intensities and orientations.}\label{tbl1}
\begin{tabular*}{\tblwidth}{@{} LLLLL@{} }
\toprule
 $\:$ & $\Delta T_{ad}^{pk}$ [0.6 T] & $\Delta T_{ad}^{pk}$ [1.0 T] & $p$ & $\Delta T_{ad}^{pk}$[0.6 T] / $\Delta T_{ad}^{pk}$[1.0 T] \\
\midrule
A: low demagnetization & 1.46 K & 2.17 K & 0.77 & 67\% \\
B: high demagnetization & 0.41 K & 0.94 K & 1.63 & 44\% \\ 
A-B & 1.05 K &  1.23 K &  - & - \\
\bottomrule
\end{tabular*}
\end{table}

As we will see in the magnetostatic simulations, these different dependencies are due to the higher impact of the demagnetizing effect at low fields, and give rise to the non-trivial field dependence of the RMCE. As we can see in figure \ref{fig:3}, the RMCE observed with 0.6 T is very similar to that observed with 1.0 T. While the field is reduced by 40\%, the maximum $\Delta T_{ad}^{rot}$ observed is 1.12 K, 96\% of what was observed for 1.0 T (1.17 K). This is consistent with what was previously seen in Gd in terms of the maximum $\Delta T_{ad}^{rot}$ \cite{Almeida2024,Pereira_2025}. Furthermore, the temperature profile of $\Delta T_{ad}^{rot}$ in our La(Fe,Mn,Si)$_{13}$H sample is more comparable between the two fields, albeit with a slight shift in temperature. Contrastingly, in Gd, the temperature profile considerably broadens as the field intensity increases \cite{Almeida2024,Pereira_2025}.

For heat pumping applications, being able to reduce the magnetic field volume while maintaining most of the useful effect is of utmost importance, since the volume of permanent magnets required for higher magnetic fields scales very quickly, and it is the most costly raw material in devices. By reducing from 1.0 T to 0.6 T, a cylindrical Halbach array with 3 cm bore (inner diameter), the volume is reduced by 70\% (see figure \ref{A3} of the annex).

\begin{figure}
    \centering
    \includegraphics[width=140mm]{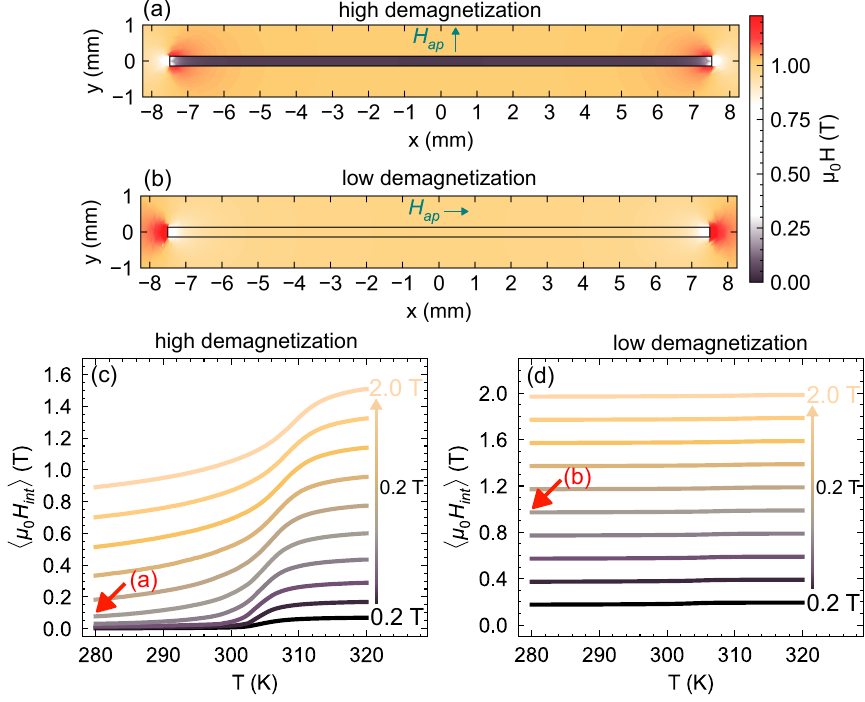}
    \caption{Cross section of the magnetic field strength in the La(Fe,Mn,Si)$_{13}$H sample and near surroundings at 280 K with 1.0 T applied field in the (a) high demagnetization, and (b) low demagnetization orientations of the applied magnetic field ($H_{ap}$). The results for all other temperature and applied field intensities for (c) the high demagnetization, and (d) low demagnetization orientations.}
    \label{fig:4}
\end{figure}

In figure \ref{fig:4}, the results of the magnetostatic simulations obtaining the average internal magnetic field strength for the experimentally measured La(Fe,Mn,Si)$_{13}$H plate shape are shown. In figure \ref{fig:4}(a) and (b), we show a cross section of the magnetic field strength distribution within the plate for (a) the high demagnetizing field orientation, when the applied field, $H_{ap}$, is perpendicular to the plate, and (b) the low demagnetizing field orientation, when $H_{ap}$ is parallel to the plate. Both the examples correspond to the same applied magnetic field strength (1.0 T) and lowest temperature (T=280 K), where the demagnetizing effect is maximized. Clearly, the internal magnetic field profile and strength are considerably different in each orientation. While the applied magnetic field is 1.0 T, the average internal magnetic field strength ($\mu_0 H$) over the entire sample volume is only 0.08 T in the high demagnetization orientation (a), while nearly identical to the applied field, 0.97 T, in the low demagnetization orientation (b). The results of the volume-average internal magnetic field strengths for all other applied field intensities and temperatures are shown for both orientations in figure \ref{fig:4}(c) and (d).

Figure \ref{fig:4}(c) shows how the average internal magnetic field in a high demagnetization orientation is considerably reduced when the magnetization of a sample increases. The temperature dependencies of the average internal magnetic field somewhat mirror the material's magnetization temperature dependency. On the other hand, in figure \ref{fig:4}(d), this dependency is barely noticeable in the low demagnetization orientation, with the internal field practically corresponding to the externally applied field.

A closer inspection of figure \ref{fig:4}(c) reveals how the demagnetizing effect has a higher relative impact in the MCE in low magnetic field intensities: while the external field is linearly increasing in 0.2 T steps between each curve, starting at 0.2 T, below T$_C$ the average internal field is very slow to increase, with the 3 curves corresponding to 0.2 T, 0.4 T, and 0.6 T practically collapsing to 0.0 T at low temperatures. 

These simulations enable the calculation of the real (taking demagnetization into consideration) $\Delta S_{iso}$ occurring in our sample for each orientation. Using equation \ref{eq:3}, for each temperature and external magnetic field intensity, we compute the integral with $H_{int,f}=\langle \mu_0 H_{int}\rangle$, where $\langle \mu_0 H_{int}\rangle$ varies with temperature and external field amplitude, as shown in figure \ref{fig:4}. To compute the rotating isothermal entropy change, $\Delta S_{iso}^{rot}$, one changes the lower limit of the integral to the initial internal field  strength, $H_{int,i}$, which is equivalent to the difference between $\Delta S_{iso}$ at each orientation:

\begin{equation}
    \label{eq:4}
    \begin{split}
           \Delta S_{iso}^{rot}(T,H_{int,f},H_{int,i}) & = & \int _{H_{int,i}} ^{H_{int,f}} \frac{\partial M(T,H_{int})}{\partial T} dH_{int} \Leftrightarrow \\
           \Delta S_{iso}^{rot}(T, H_{int,f},H_{int,i}) &=& \Delta S_{iso}(T,H_{int,f})-\Delta S_{iso}(T,H_{int,i}).
    \end{split}
\end{equation}

\begin{figure}
    \centering
    \includegraphics[width=160mm]{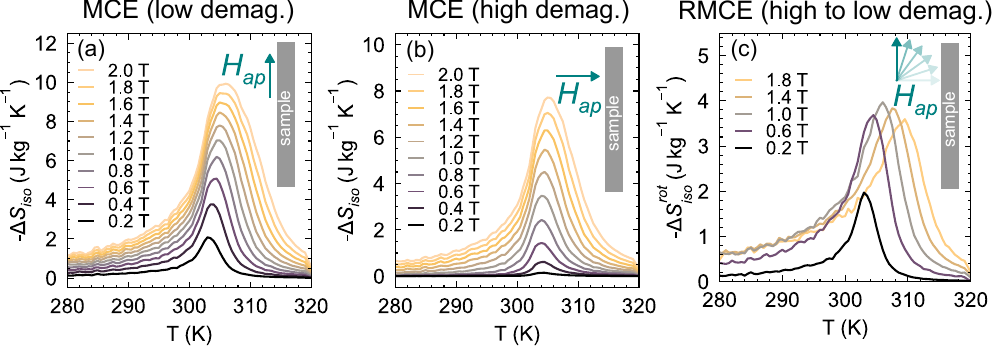}
    \caption{Isothermal entropy change, $\Delta S_{iso}$ as obtained through the Maxwell relation in conjunction with the results of the magnetostatic simulations for the (a) low demagnetization and (b) high demagnetization orientations of the applied magnetic field, $H_{ap}$, as shown in the inset schematic. (c) Rotating isothermal entropy change, $\Delta S_{iso}^{rot}$, occurring upon rotating $H_{ap}$ from the high to the low demagnetization orientation,  corresponding to the difference between $\Delta S_{iso}$ of each orientation. }
    \label{fig:5}
\end{figure}

In figure \ref{fig:5}, the results for $\Delta S_{iso}$ and  $\Delta S_{iso}^{rot}$ up to 2.0 T applied magnetic field intensity are shown. The maximum $\Delta S_{iso}$ obtained in the low demagnetization orientation at 1.0 T is 7.04 J kg$^{-1}$ K$^{-1}$. Similarly to what was observed in the $\Delta T_{ad}$ measurements, this value is considerably less than previously observed in a sample of equivalent composition, 14.14 J kg$^{-1}$ K$^{-1}$ \cite{Basso2015}. This  50\% discrepancy to the reference is clearly related to the large (25 wt\%) content of $\alpha$-Fe in our sample, but too large to be the only cause. Another contributing factor may be the difference in the method used for reference measurements, which is based on calorimetry rather than magnetometry. Another study, performing magnetization measurements on an equivalent sample, obtained a maximum $\Delta S_{iso}$ of 11.92 J kg$^{-1}$ K$^{-1}$ for 1.6 T \cite{Barcza2011}. Our maximum $\Delta S_{iso}$ for 1.6 T in the low demagnetization orientation is 8.97 J kg$^{-1}$ K$^{-1}$ value, 25\% less than this reference, which is consistent with what was found for the maximum observed $\Delta T_{ad}$, suggesting the high amount of secondary $\alpha$-Fe phase may indeed be the cause for the attenuated values.

As observed with the $\Delta T_{ad}$ measurements, it is useful to discuss the exponent $n$ of the field dependence of the maximum $\Delta S_{iso}$. In the low demagnetization orientation (as shown in the inset of figure $\ref{fig:2}$ for no demagnetizing factor) the exponent decreases slightly with field increase, closely following the polynomial dependence with $p=0.70$ up to 1.0 T ($R^2=0.997$), and reducing to $p=0.51$ ($R^2=0.990$) for higher fields. On the other hand, for the high demagnetization orientation, $\Delta S_{iso}$ is much slower to increase, with $p=1.5$ up to 1.0 T ($R^2=0.998$).  

This accelerated dependence of $\Delta S_{iso}$ observed for high demagnetizing factors has been discussed previously \cite{Romero-Muñiz2014}, and provides an intuitive explanation for the non-trivial magnetic field dependence of $\Delta S_{iso}^{rot}$, which is shown in figure \ref{fig:5}(c). The profiles and relative amplitudes of $\Delta S_{iso}^{rot}$ for 0.6 T and 1.0 T are remarkably similar to the experimentally measured $\Delta T_{ad}^{rot}$ curves for the respective fields. Additionally, we see that $\Delta S_{iso}^{rot}$ persists on gradually drifting towards higher fields, as shown for 1.4 T and 1.8 T, whose respective $\Delta T_{ad}^{rot}$ curves were not experimentally achievable to measure in the present setup.

Finally, we underline that despite the aforementioned high wt\% of secondary $\alpha$-Fe phase in this sample, 3.68 J K$^{-1}$ kg$^{-1}$ maximum $\Delta S_{iso}^{rot}$ were obtained for 0.6 T, which is +194\% of what has been observed in Gd for the same magnetic field \cite{Almeida2024}. Indeed, it is comparable to the expected maximum of the \textit{conventional} MCE ($\Delta S_{iso}$) in Gd for 1.0 T, (between 3-4 J K$^{-1}$ kg$^{-1}$ \cite{Lyubina2011,Smith2012}). At the same time, it should be noted that the maximum $\Delta T_{ad}^{rot}$ being considerably lower than the maximum $\Delta T_{ad}$ (conventional MCE) of Gd is not a trivial hindrance to implementation in devices \cite{Klinar2024}. A fair comparison from the point of view of applications should adopt a holistic perspective, taking into account material properties as well as the advantages and disadvantages of the RMCE to the MCE previously mentioned.

\section{Conclusions and outlook}

The conventional and rotating magnetocaloric effects of a fully hydrogenated La(Fe,Mn,Si)$_{13}$ sample produced by powder metallurgy were characterized. Due to the sample's high aspect ratio ($\sim$50) shape, the MCE is considerably different depending on the orientation of the applied magnetic field due to the demagnetizing effect. 

Despite the relatively high $\alpha$-Fe content of the sample (21-27\%), a considerable maximum $\Delta T_{ad}$ of 2.17 K was found when applying a 1.0 T magnetic field in the low demagnetization orientation, which is reduced to 0.94 K if the magnetic field is applied in the high demagnetization orientation. 

The maximum $\Delta T_{ad}^{rot}$ observed was 1.17 K, for 1.0 T. Reducing the magnetic field intensity by 40\%, to 0.6 T, resulted in a maximum $\Delta T_{ad}^{rot}$ of 1.12 K, 96\% of that obtained with 1.0 T, corresponding to a reduction of 70\% permanent magnet volume. 

Magnetostatic simulations enabled the calculation of $\Delta S_{iso}^{rot}$ for the experimentally measured sample shape, yielding remarkably similar profiles to those of $\Delta T_{ad}^{rot}$ observed experimentally for 0.6 T and 1.0 T. 

The maximum $\Delta S_{iso}^{rot}$ obtained was 3.97 J K$^{-1}$ kg$^{-1}$, for 1.0 T. Similarly to what was observed for $\Delta T_{ad}^{rot}$, the maximum $\Delta S_{iso}^{rot}$ observed for 0.6 T was 3.68 J K$^{-1}$ kg$^{-1}$, 7\% less than with 1.0 T, despite the 40\% reduction in applied field intensity. These maximum $\Delta S_{iso}^{rot}$ is comparable to what is achievable with the conventional MCE in Gd with a magnetic field of 1.0 T. This highlights the potential of the RMCE for magnetic refrigeration applications using low magnetic field intensities. At the same time, the relatively low maximum $\Delta T_{ad}^{rot}$ (<2 K) may be a challenge to device performance, which still needs to be assessed holistically.

\section*{Acknowledgements}

The authors acknowledge FCT and its support through the projects LA/P/0095/2020, UIDB/04968/2025, and UIDP/04968/2025. R. Almeida thanks FCT for his PhD grant with reference 2022.13354.BD and DOI \\ https://doi.org/10.54499/2022.13354.BD. This project has received funding from the European Union’s Horizon Europe research and innovation programme through the European Innovation Council under the grant agreement No. 101161135 – MAGCCINE. This work was also developed within the scope of the project CICECO Aveiro Institute of Materials, UID/50011/2025 (DOI 10.54499/UID/50011/2025) \& LA/P/0006/2020 (DOI 10.54499/LA/P/0006/2020), financed by national funds through the FCT/MCTES (PIDDAC).

The authors also acknowledge Rui Rocha from CEMUP Lab for his contribution to SEM analyses and interpretation. 

\appendix

\section{Phase fraction from bulk magnetometry}

\begin{figure}
    \centering
    \includegraphics[width=70mm]{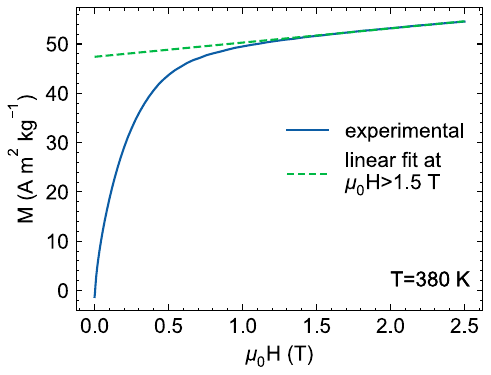}
    \caption{Magnetization versus temperature response of the La(Fe,Mn,Si)$_13$H sample at 380 K, well above its phase transition temperature. A linear fit, assuming linear susceptibility, allows a rough estimation of the ferromagnetic $\alpha$-Fe phase content.}
    \label{A1}
\end{figure}

Assuming a constant susceptibility (linear dependence) of the paramagnetic 1-13 phase, we may estimate the the weight percentage of $\alpha$-Fe in our sample by isolating its ferromagnetic contribution. A linear fit disregarding the region below 1.5 T applied magnetic field ($R^2$=0.998) results in a y-intersect of 47.40 A m$^2$ kg$^{-1}$ (see figure \ref{A1}). Assuming the saturation moment of Fe, 222 A m$^2$ kg$^{-1}$ \cite{Danan1968}, this corresponds to 21.4 wt.\% of the sample.

\section{Phase fraction from SEM}

SEM imaging of this alloy has been shown some times before \cite{ROSENDAHLHANSEN20103447,WANG2022164274, WANG2023134034}. The three main phases are clearly differentiated through backscattered electron imaging.

\begin{figure}
    \centering
    \includegraphics[width=140mm]{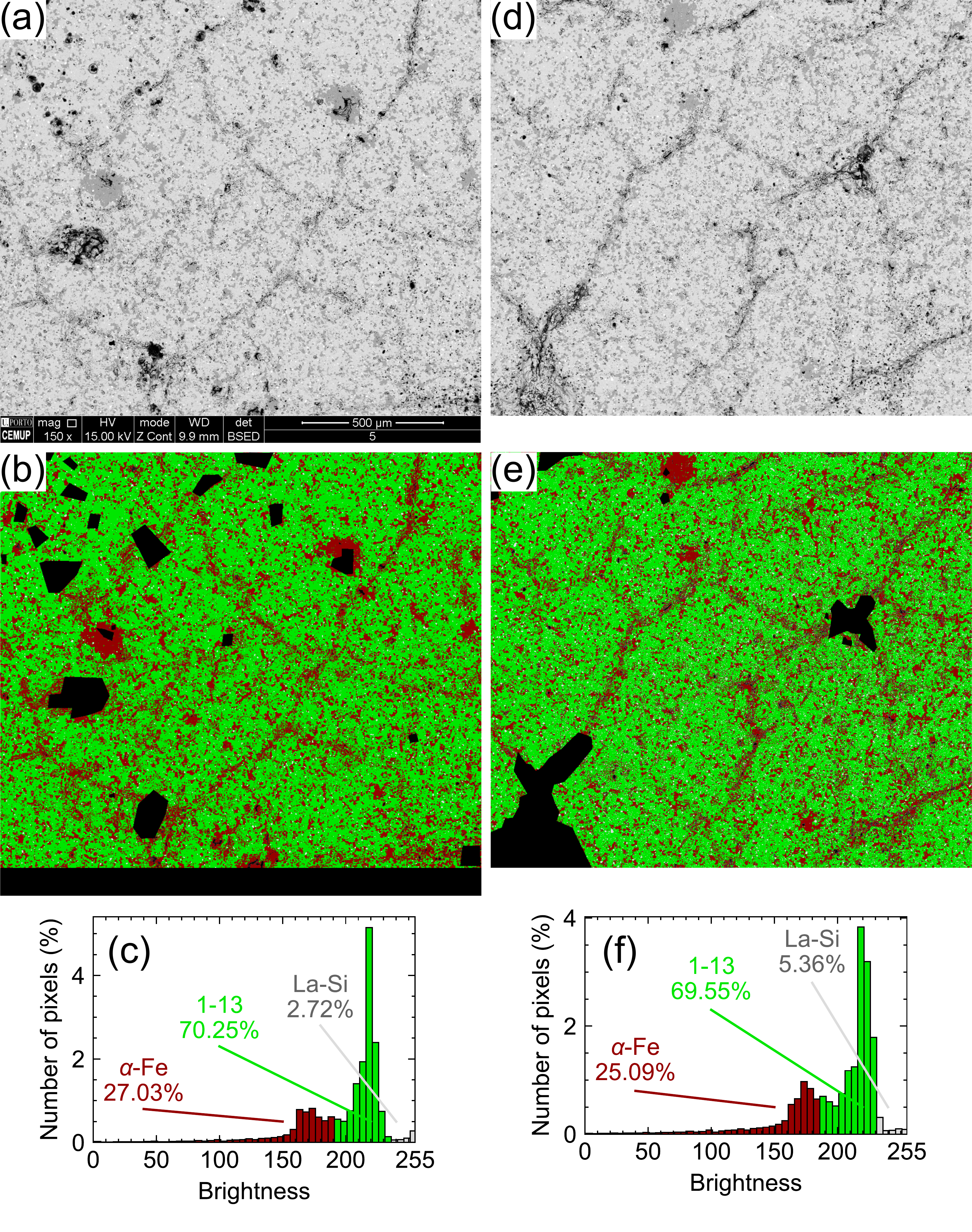}
    \caption{Backscattered electron SEM imaging showing (a) a representative image of the surface of our sample, (b) the three phases separated by filtering through ranges in brightness, and (c) a histogram of the pixel brightnesses and the corresponding phases. The same is shown for a second representative image of a different section in the face of the sample (d-f). Before processing, areas with a high density of defects are replaced with black pixels, which are not considered for phase fraction determination.}
    \label{A2}
\end{figure}

Figure \ref{A2} shows two representative images of the surface of our sample and their analyses. Due to imperfect polishing, there are some regions with considerable defects which were replaced with black pixels and thus not included for phase fraction determination. To improve confidence, 6 other images of the same sample were analysed similarly, which yielded the final phase fraction estimations: 69.5\% main 1-13 phase, 28.0\% secondary $\alpha$-Fe phase, and 2.5\% La-rich.

\section{Cylindrical Halbach array magnetic field and total volume}

\begin{figure}
    \centering
    \includegraphics[width=70mm]{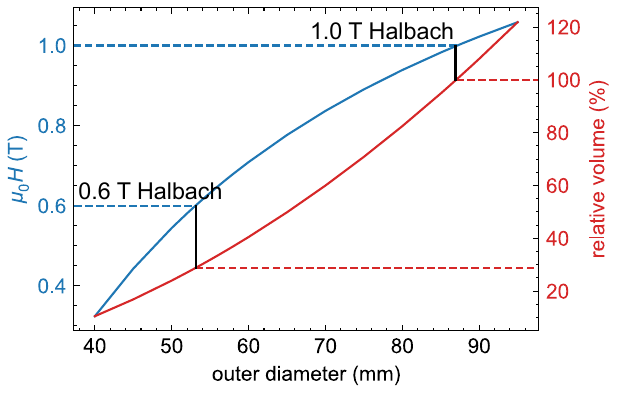}
    \caption{Magnetic field produced (y-axis on the left) and relative total volume (y-axis on the right) of a Halbach cylinder for varying outer diameter, with fixed inner diameter of 3 cm and length of 4 cm.}
    \label{A3}
\end{figure}

To give a sense of how much more permanent magnet material one requires for a certain increase of magnetic field (or, conversely, how much one saves by not requiring a certain high value), we obtained the resulting magnetic field supplied by a Nd-Fe-B Halbach cylinder with an inner bore of 3 cm and a varying outer diameter. Considering the Halbach supplying 1.0 T as our reference (100\%), we obtain that for supplying 0.6 T one reduces the total volume of permanent magnet material needed by 71\%. This is shown in figure \ref{A3}.

\printcredits

\bibliographystyle{ieeetr}

\bibliography{bib}

\end{document}